\input harvmac

\Title{HUTP-97/A008}{Cosmic Ray and Neutrino Tests of  Special Relativity}
\centerline{ Sidney  Coleman \& Sheldon L. Glashow }\bigskip
\centerline{Lyman Laboratory of Physics}
\centerline{Harvard University}\centerline{Cambridge, MA 02138}
\vskip .3in

\noindent Searches for anisotropies due to Earth's motion relative to a
preferred frame --- modern versions of the  Michelson-Morley experiment ---
provide precise verifications of special relativity. We describe other
tests, independent of this motion, that are or can become even more
sensitive. The existence of high-energy cosmic rays places strong
constraints on  Lorentz non-invariance. Furthermore, if the maximum
attainable speed of a particle depends on its identity, then neutrinos,
even if massless, may exhibit flavor oscillations. Velocity differences far
smaller than any previously probed can produce characteristic effects at
accelerators and solar neutrino experiments.

\Date{04/97}

Is the special theory of relativity,  for reasons unspecified and unknown, 
only an approximate symmetry of nature?   To investigate possible
violations of Lorentz symmetry, we follow earlier analyses \ref\rwill{C. M.
Will, {\it Theory and Experiment in Gravitational Physics,} (Cambridge
University Press, 1993); \ See also: M. P. Haugan and C. M. Will, Phys.
Today 40 (May, 1987) 69; \ E. Fischbach, M. P. Haugan, D. Tadic, and H.-Y.
Cheng, Phys. Rev. D32 (1985) 154; \ G. L. Greene, M. S. Dewey, E. G.
Kessler, Jr., and E. Fischbach, Phys. Rev. D44 (1991) 2216. L.
Gonzales-Mestres (to be publ.) discusses possible Lorentz non-invariance in
a  different context.} by assuming
the laws of physics to be invariant under rotations and translations in a
preferred reference frame $\cal F$. 
This frame is often taken to be the `rest frame
of the universe,' the frame in which
the cosmic microwave background is isotropic.  To
parameterize departures from Lorentz invariance,  standard practice has
been to modify Maxwell's equations while leaving other physical laws
intact.

Although we shall shortly consider more general Lorentz non-invariant
perturbations, let us for the moment adhere to standard practice:  we
assume that the only Lorentz non-invariant term in $\cal L$ is proportional
to the square of the magnetic field strength.  Thus, the {\it in vacua\/}
speed of light $c$ differs from the maximum attainable speed of a material
body (here taken to be unity). The small parameter $1-c$ completely
characterizes this departure from special relativity in $\cal F$. In a
frame moving at velocity $\vec u$ relative to $\cal F$,  the velocity of 
light $c^\prime$ depends on its angle $\theta$  relative to $\vec u$. For
$u\ll 1$, we find $c^\prime(\theta)\simeq  c+2(c-1)u\cos{\theta}$.  The
failure of rotational invariance in the laboratory frame  leads to
potentially observable effects that are 
proportional to $u^2(1-c^2)$. Searches for
these anisotropies yielding null results have provided  precision tests of
special relativity.

A laser-interferometric Michelson-Morley experiment \ref\rbr{A.
Brillet and J. L. Hall, Phys. Rev. Lett. 42 (1979) 549.}  found $\vert
1-c\vert<10^{-9}$. Atomic physicists obtained
stronger constraints 
using techniques pioneered by Hughes and Drever \ref\rhd{V. W. Hughes, H.
G. Robinson, and V. Beltran-Lopez, Phys. Rev. Lett. 4 (1960) 342; \  R. W.
P. Drever, Philos. Mag. 6 (1961) 683.}. Prestage {\it et al.} \ref\rpres{J.
D. Prestage, J. J. Bollinger, W. M. Itano, and D. J. Wineland,  Phys. Rev.
Lett. 54 (1985) 2387.} found $<10^{-18}$ and  Lamoreaux {\it et al.} 
\ref\rlam{S. K. Lamoreaux, J. P. Jacobs, B. R. Heckel, F. J. Raab, and E.
N. Fortson, Phys. Rev. Lett. 57 (1986) 3125.} set the current limit on the
velocity difference, 
\eqn\eone{\vert 1-c\vert  <3\times 10^{-22}\,.}  
These limits are obtained for  $\cal F$ 
at rest relative to the cosmic background radiation, whence
$u\simeq 10^{-3}$.
They  would be two orders of magnitude weaker were  $\cal F$
at rest relative to the Sun.  

We find  additional limits on $1-c$ that do not require precision
experiments, yet are comparable in sensitivity to \eone. For $u\ll 1$, the
new constraints do not depend on the motion of the laboratory relative to
$\cal F$.  They follow from the mere existence of high-energy cosmic rays.
Suppose $c>1$. Because the photon 4-momentum $(E/c,\,E)$   is timelike, a
sufficiently energetic photon can and will decay rapidly into an
electron-positron pair.  The threshold energy for $\gamma\rightarrow
e^++e^-$ is $E_0=2m/\sqrt{c^2-1}$, with $m$ the electron mass.  In
first-order perturbation theory (using an invariant matrix element and the
modified photon dispersion relation), we obtain for the decay rate in $\cal
F$:
\eqn\eT{\Gamma = \alpha (c-1)E \left\{1-(E_0/E)^2\right\}^{3/2}\,.}
A photon that is produced with energy well above $E_0$ decays
rapidly, with mean lifetime
$\tau\simeq 3 (E_0/E)/\sqrt{c-1}$~ns. Hence
a primary cosmic-ray photon with
$E>E_0$ cannot reach Earth. However, primary photons with energies up to
20~TeV have been
seen \ref\rcro{J. W. Cronin,
K. G. Gibbs, and T. C. Weekes, Annu. Rev. Nucl. Part. Sci. 43 (1993) 883.}.
Thus we obtain the bound:\footnote{$^1$}{If 
$\cal F$ moves relativistically ({\it i.e.,} if $u\sim 1$),
our arguments change but our conclusion remains. 
Lorentz non-invariance would be signaled by a pronounced (and unseen) 
dipole anisotropy of the highest energy cosmic rays.}
\eqn\enewa{c-1 < 1.5\times 10^{-15}\,.}
(If this bound were saturated, the threshold energy for photon decay would
be 18.6~TeV and the mean range of a 20~TeV photon would be 8~cm.)
Eq. \enewa\ is weaker than
\eone, but it arises, so to speak, for free.

Suppose $c<1$. A charged particle traveling faster than light 
loses energy rapidly
via vacuum \v Cerenkov radiation. The threshold 
energy for $p\rightarrow p+\gamma$
is $E^\prime_0 = M/\sqrt{1-c^2}$, with $M$ the 
particle mass. In first-order perturbation theory, the
rate of energy loss in $\cal F$ is:
\eqn\ededx{{dE\over dx} \simeq -{\textstyle{1\over 3}}\alpha Z^2E^2
(1-c)\left\{1-(E^\prime_0/E)^2\right\}^3
\left\{1-{\textstyle{3\over 8}}(E^\prime_0/E)^2\right\}\,.}
A proton produced with energy well above $E^\prime_0$ radiates photons copiously
until its energy approaches $E^\prime_0$. 
It follows that a primary cosmic-ray 
proton with  $E>E^\prime_0$ cannot reach Earth. 
However, primary protons with energies up to
$10^{20}$~eV have been seen \ref\rquery{J. Wdowczyk
and A. W. Wolfendale, Annu. Rev. Nucl. Part. Sci. 39 (1989) 43.}. We
thereby obtain the bound:$^1$ 
\eqn\enewb{1-c < 5\times 10^{-23}\,.}
(If this bound were saturated, the vacuum \v Cerenkov threshold for a
proton would be
$E_0'=9.38\times 10^{19}$~eV.
One traversing empty space with  any much greater energy
would be reduced to an energy of $1.1\times E_0'$ less
than 140~cm from its point of production.)
Eq. \enewb\ is considerably more restrictive than \eone.

From a field-theory viewpoint, there is no reason to restrict ourselves to
just the effects of a tiny $B^2$ term.  Properly, we should begin with the
conventional Lagrangian $\cal L$ of the standard model of particle physics
and introduce all renormalizable Lorentz non-invariant interactions
consistent with our symmetry assumptions: rotational and 
translational invariance in the preferred frame, and  the
$SU(3)\times SU(2)\times U(1)$
gauge symmetry of the standard
model.
(The condition of renormalizability, that the dimensions of the non-invariant
interactions be no greater than four, arises if we assume the
fundamental source of non-invariance to occur at a very large mass scale.
Then higher-dimension operators are supressed by inverse powers of the large
mass.)  We are preparing a study \ref\rsc{Sidney Coleman and Sheldon L.
Glashow, (manuscript in preparation).} of the general case; in this note we
restrict ourselves to a few observations.

Of the many conceivable  Lorentz non-invariant additions to the Lagrangian,
some are TCP even, the others odd.   Aside from $B^2$, the only 
renormalizable term that affects the propagation of light is the TCP odd
expression:
\eqn\etcp{ {\textstyle{1\over 2}}\,
\epsilon^{\mu\nu\lambda\sigma}\,F_{\mu\nu}\,A_\lambda\,s_\sigma
\equiv 
{\vec A}\cdot {\vec B}\,s_0 + ({\vec A}\times{\vec E})\cdot
{\vec s}\;,}
where $s_\mu$ is a fixed 4-vector with dimension of reciprocal length.
 This term is not gauge
invariant, but it makes a gauge-invariant contribution to the action
\ref\rcfj{M. Carroll, G.B. Field and R. Jackiw, Phys. Rev. D40 (1990)
1231.}. It causes the linear polarization of a photon that travels a
distance $r$ to
rotate by the angle 
$\beta = (s_0 + {\vec s}\cdot {\hat n})r$, where $\hat n$ is a unit vector
along the photon direction. 
Analyses of astronomical data 
\rcfj\ \ref\rgt{Maurice
Goldhaber and Virginia Trimble, J. Astrophys. Astr. 17 (1996) 17.}
severely constrain the rotationally invariant term in \etcp.  They 
lead to the constraint $s_0<10^{-28}$~cm$^{-1}$. More recently, 
indications of an anisotropy
of electromagnetic propagation at cosmological distances
\ref\rnod{B. Nodland and J.P. Ralston, Phys. Rev. Lett. 78 (1997) 3043}
have been reported. This effect
can be interpreted in terms of \etcp\ with
$\vert{\vec s}\vert\simeq 
10^{-27}$~cm$^{-1}$. (Accepted at face value, these results
would show that $s$ is space-like, and that there is no frame with
rotational symmetry.)

The observations discussed above
make the detection of any TCP violating effects in the
microworld unlikely.  Radiative corrections induced by significant TCP
violation elsewhere in $\cal L$ would be expected to induce the term
$\etcp$, and conversely.  Thus, the largest dimensionless 4-vector that
might characterize TCP violation in particle physics is
$s_\mu/m$ (with $m$ the electron mass). Its components have been shown to
be less than $4\times 10^{-38}$, far smaller than any of the limits we have
discussed.  For this reason,  we shall examine only those violations of
Lorentz invariance conserving   TCP.

Tiny Lorentz non-invariant (but TCP conserving) additions to the matter
portion of  the Lagrangian affect the free propagation of a particle in a
fashion depending on its identity and helicity \rsc. In particular,
neutrinos may differ in their maximum attainable 
 velocities. Massless 
neutrinos  cannot oscillate if special relativity is unbroken. However,
they can oscillate if different neutrinos travel at  slightly different
speeds {\it in vacua.} 

Let
$\nu_i$ denote the velocity eigenstates of neutrinos with speeds
$1+ v_i$, where $\vert v_i\vert
\ll 1$. These states
may not coincide with the flavor eigenstates set by weak
interactions. Thus,
a massless neutrino produced in one flavor state can appear in
another flavor state along its way. 
Consider oscillations
between the two states:
\eqn\etwo{\nu_\mu=\cos{\theta_v}\,\nu_1 + \sin{\theta_v}\,\nu_2\,, \qquad
\nu_e =\cos{\theta_v}\,\nu_2 - \sin{\theta_v}\,\nu_1\,.}
A newly-born muon neutrino with definite momentum propagates 
through empty space as a
linear combination of states with slightly different energies:
$E_1-E_2 \simeq \delta v \,E$, where $\delta v= v_1-v_2$. The probability
for it to be an electron neutrino after traversing a distance $R$ is:
\eqn\ethr{P = \sin^2{2\theta_v}\, \sin^2\left\{\delta vER/2\right\}\,.}
%
A result similar to \ethr\ was obtained 
\ref\rvep{M. Gasperini, Phys. Rev. D38 (1988) 2635; A. Halprin and C.N.
Leung, Phys. Rev. Lett. 67 (1991) 1833.} in the context of 
departures from the equivalence principle rather than special relativity.
For these (VEP) oscillations, $\delta v$ is replaced by a difference
of terms proportional
to the local gravitational potential, with tiny coefficients depending on
neutrino identity.
The energy dependence of
`velocity oscillations' differs from that of
conventional `mass oscillations,' for which
\eqn\efou{P \simeq 
\sin^2{2\theta_m}\, \sin^2\left\{\delta m^2 R/4E\right\} \,.}
%

Observations of neutrinos emitted from supernova SN1987a limit
possible Lorentz non-invariance in the neutrino sector. 
From reference \ref\rssb{A. Y. Smirnov, D.N. Spergel and J.N. Bahcall,
Phys. Rev. D49 (1994) 1369.}, we find $\sin^2{2\theta_v}< 0.35$ for
$\delta v> 10^{-35}$. Furthermore, the velocity of $\nu_e$ ({\it i.e.,}
its dominant
velocity eigenstate) cannot differ from the velocity of light by more than two
parts in $10^9$ \ref\rleo{M.J. Longo, Phys. Rev. D36 (1987) 3276\semi
L. Stodolsky, Phys. Lett.  B201 (1988) 353.}.

Accelerator and reactor
experiments constraining the mass-oscillation parameters $\delta m^2$
and $\theta_m$ must be reanalyzed to yield constraints on the
velocity-oscillation parameters $\delta v$ and $\theta_v$. For the case of
$\nu_\mu\rightarrow \nu_e$ with  maximal mixing, a cursory examination
of the data yields the tentative estimate:\footnote{$^2$}{Our result follows
from the high-energy experiment of
Vilain {\it et al.} \ref\rvil{P. Vilain {\it et al.,} Z. Phys. C64
(1994) 539.}.  The limit for mass oscillations,
$\delta m^2 < 0.09$~eV$^2$, arises from the
lower energy experiment of C. Angelini {\it et al.} \ref\rang{C. Angelini
{\it et al.,} Phys. Lett. B179 (1986) 307.}.}
\eqn\eest{\vert v_1-v_2 \vert  < 10^{-21}\,,}
a result comparable to the analogous atomic-physics limit
\eone. Much of the literature concerned with VEP oscillations
\ref\rjb{{\it E.g.,} J.N. Bahcall, P.I. Krastev and C. N. Leung,
Phys. Rev. D52 (1995) 1770\semi
 A. Halprin, C.N. Leung, and J. Pantaleone, Phys. Rev. D53 (1996) 5365;
S.L Glashow, A. Halprin, P.L. Krastev,
C.N. Leung, and J. Pantaleone (to be published).}
is applicable to
the phenomenology of velocity oscillations. 
Neutrino experiments at existing accelerators, with longer
baselines and higher energies, can search for velocity differences
among neutrinos hundreds of times smaller than the current upper
bound.

Lorentz non-invariant velocity oscillations,
like VEP oscillations \rjb, can affect the solar neutrino
flux. Suppose neutrinos were massless.
If their velocity mixing were maximal and $\delta v$
were $8.3\times 10^{-24}$, all 1~MeV solar electron neutrinos would arrive
at Earth as muon neutrinos. This
velocity difference is two powers of ten
less than the upper bounds \eone\ or  \eest. 
Their characteristic energy dependence should enable
experimenters to distinguish vacuum velocity oscillations from other proposed
solutions to the solar neutrino problem, and to place stringent new limits
on the associated departures from Lorentz invariance.

What if neutrinos do have mass? In this case, there are three possibly
distinct bases for their description: flavor eigenstates (set by weak
interactions), mass eigenstates (energy eigenstates at zero momentum), and
velocity eigenstates (energy eigenstates at infinite
momentum).
Massive neutrinos may undergo simultaneous velocity and mass
oscillations. For a two-state system, the probability for identity
change is given \rsc\ by a  formula much like \ethr\ or
\efou:
\eqn\efiv{P= \sin^2{2\Theta}\,\sin^2\left\{
\Delta\,R/4E\right\}\,,}
where the mixing angle $\Theta$ 
and phase factor $\Delta$ are:
\eqn\esix{\eqalign{
\Delta\sin{2\Theta}\,&=\left\vert\delta m^2\sin{2\theta_m}+2E^2\delta v
\sin{2\theta_v}\right\vert\,,\cr
\Delta\cos{2\Theta}\,&=\left\vert\delta m^2\cos{2\theta_m}+
2e^{i\phi}E^2\delta v
\cos{2\theta_v}\right\vert\,.\cr}}
The result depends on $E$, and as well, on  $\delta m^2$,
$\delta v$, $\cos{\theta_m}$, and $\cos{\theta_v}$. An additional
parameter $\phi$ appears because velocity and mass eigenstates, in general,
are
related by a complex unitary transformation.
Eq.\efiv\ reduces to mass mixing \efou\ for small $E$, and to 
velocity mixing \ethr\  for large $E$.
If all three neutrinos are both mass and velocity mixed,
the analysis is more complicated
and neutrino experimenters would be presented
with an intricate challenge.

Searches failing to detect neutrino velocity oscillations can
provide new and more sensitive tests of special relativity. Successful
searches would reveal a surprising connection between cosmology (which
has a preferred reference frame) and particle physics (which 
ordinarily does not).

\bigbreak\bigskip\centerline{\bf Acknowledgement}\nobreak
One of the authors (SLG) is grateful for illuminating discussions with,
or communications from: John Bahcall,
Jerry Gabrielse, Maurice Goldhaber, 
John Learned, Sandip Pakvasa, Georg Raffelt, and Xerxes Tata.
This work was supported in part
by the National Science Foundation under
grant NSF-PHYS-92-18167.

\listrefs
\bye